\documentclass[reprint,showpacs,amsfonts,amssymb,amsmath,aps,pra,
               superscriptaddress]{revtex4-1}

\usepackage[T1]{fontenc}
\usepackage{microtype}
\usepackage{graphicx}
\usepackage{microtype}
\usepackage[pdftex,colorlinks]{hyperref}
\hypersetup{%
  linkcolor=blue,%
  citecolor=blue,%
  urlcolor=blue,%
  pdftitle={Vortices in Bose-Einstein condensates with PT-symmetric gain and loss},%
  pdfauthor={Lukas Schwarz, Holger Cartarius, Ziad H. Musslimani, J\"org Main, G\"unter Wunner}%
}

\usepackage{xspace}
\usepackage{braket}
\newcommand{\PT}{$\mathcal{PT}$\xspace}

\newcommand{\e}{\mathrm{e}}
\newcommand{\im}{\mathrm{i}}
\renewcommand{\Im}{\operatorname{\mathsf{Im}}}

\begin{document}
\title{Vortices in Bose-Einstein condensates with \PT-symmetric gain and loss}
\author{Lukas Schwarz}
\affiliation{Institut f\"ur Theoretische Physik 1, Universit\"at Stuttgart, Pfaffenwaldring 57, 70550 Stuttgart, Germany}
\author{Holger Cartarius}
\affiliation{Institut f\"ur Theoretische Physik 1, Universit\"at Stuttgart, Pfaffenwaldring 57, 70550 Stuttgart, Germany}
\author{Ziad H. Musslimani}
\thanks{Part of this work was done during ZHM's research stay at the University
  of Stuttgart. Valuable support from the German Academic Exchange Service
  (DAAD) is gratefully acknowledged.}
\affiliation{Institut f\"ur Theoretische Physik 1, Universit\"at Stuttgart, Pfaffenwaldring 57, 70550 Stuttgart, Germany}
\affiliation{Department of Mathematics, Florida State University, Tallahassee, Florida 32306-4510, USA}
\author{J\"org Main}
\affiliation{Institut f\"ur Theoretische Physik 1, Universit\"at Stuttgart, Pfaffenwaldring 57, 70550 Stuttgart, Germany}
\author{G\"unter Wunner}
\affiliation{Institut f\"ur Theoretische Physik 1, Universit\"at Stuttgart, Pfaffenwaldring 57, 70550 Stuttgart, Germany}

\date{\today}
\begin{abstract}
  We investigate vortex excitations in dilute Bose-Einstein condensates in
  the presence of complex \PT-symmetric potentials. These complex potentials
  are used to describe a balanced gain and loss of particles and allow for
  an easier calculation of stationary states in open systems than in a full
  dynamical calculation including the whole environment. We examine the
  conditions under which stationary vortex states can exist and consider
  transitions from vortex to non-vortex states. In addition, we study the
  influences of \PT symmetry on the dynamics of non-stationary vortex
  states placed at off-center positions.
\end{abstract}
\maketitle

\section{Introduction}
Quantized vortices \cite{RevModPhys.71.463,JPhysCondensMatter.13.R135,ModPhysLettB.18.1481} are coherent excitations in Bose-Einstein condensates, which
can arise due to the superfluid property of such a condensate. These are stationary flows characterized by their quantized circulation
\begin{align}
    \Gamma = \oint \vec v \cdot \mathrm d\vec r = \oint \frac \hbar m \nabla S \cdot \mathrm d\vec r = \frac{2\pi\hbar}{m} n\,, \qquad n \in \mathbb{N}\,,
\end{align}
where $\vec v = \vec j/\rho$ is the probability current velocity, $\vec j$ the probability current density, $\rho$ the probability density, and $S$ the phase of the wave function $\psi = \sqrt{\rho}\e^{\im S}$. Vortices in Bose-Einstein condensates have been studied theoretically \cite{SovPhysJETP.13.451,IlNuovoCimento.20.454,JMathPhys.4.195} since the introduction of mean-field theory for Bose-Einstein condensates. Later on, several techniques to create vortices experimentally have been proposed and also realized like rotating traps \cite{PhysRev.138.A429,EurPhysJD.7.399,RevModPhys.81.647}, stirring condensates via laser beams \cite{PhysRevLett.83.2498,PhysRevLett.83.895}, phase imprinting techniques \cite{PhysRevA.60.R3381,PhysRevLett.89.190403} or temperature quenching via the Kibble-Zurek mechanism \cite{Nature.455.948,EurPhysJ.224.577}.

In this paper vortices are investigated in the presence of coherent balanced gain and loss of particles described via complex \PT-symmetric potentials. The \PT-symmetric formulation of quantum mechanics was introduced by Bender and Boettcher in 1998 \cite{PhysRevLett.80.5243}, where they found that the requirement of a Hermitian Hamiltonian is not the most general condition to ensure real energy eigenvalues required for a physical theory. It can be shown \cite{ContempPhys.46.277} that a Hamiltonian which possesses an unbroken \PT symmetry has purely real eigenvalues. Here, the unbroken \PT symmetry is defined in such a way that $[H,\mathcal{PT}] = 0$ and in addition all eigenfunctions of the Hamiltonian are simultaneously eigenfunctions of the \PT operator. Such a Hamiltonian is called pseudo-Hermitian \cite{JMathPhys.43.205,JMathPhys.43.2814,JMathPhys.43.3944} and allows for complex potentials if the real and imaginary parts are symmetric and antisymmetric functions of the position vector $\vec r$, respectively.

An imaginary part of a potential is generally interpreted as an additional gain-loss term. This can be understood if one calculates the continuity equation for the (non)linear Schr\"odinger equation with a complex potential. After a short calculation one obtains
\begin{align}
    \partial \rho + \nabla \vec j = 2\rho \Im V\,.
\end{align}
The imaginary part of the potential acts as an additional source or drain term in the range where it overlaps with the wave function. A \PT-symmetric situation differs from a general complex potential in such a way that the gain and loss contributions are balanced. This allows for stationary states with real energies, and \PT symmetry is used as an effective theory for an easier description of this dynamical process within the framework of stationary equations.

Since the discovery of Bender and Boettcher \PT symmetry has become important in
many areas of physics. Most theoretical studies are in the fields of quantum
mechanics \cite{JMathPhys.40.2201,PhysLettA.264.108,JPhysA.43.145301,%
  JPhysA.43.055307} or quantum field theories \cite{PhysRevD.85.085001,%
  FortschrPhys.61.140}. Nonlinear \PT-symmetric quantum systems have been
discussed for Bose-Einstein condensates described in a two-mode approximation
\cite{JPhysA.41.255206,JPhysA.45.444015}, in position space
\cite{FortschrPhys.61.124,JPhysA.48.335301,IntJTheorPhys.54.4054}, and in
model potentials \cite{JPhysA.41.244019}. However, the concept of \PT symmetry
could also be studied in microwave cavities \cite{PhysRevLett.108.024101} and
electronic devices \cite{PhysRevA.84.040101}.

In optics, the idea of \PT symmetry can be applied as well to describe wave guides with complex refractive indices \cite{PhysRevLett.101.080402,PhysRevLett.100.103904,PhysRevLett.100.030402,OptLett.32.2632}, and it is possible to observe this experimentally \cite{PhysRevLett.103.093902,NatPhys.6.192,NatPhys.10.394}. By contrast, an experimental confirmation of \PT symmetry in a quantum mechanical system is still lacking. However, if \PT symmetry is applied to the nonlinear Gross-Pitaevskii equation to describe a dilute Bose-Einstein condensate \cite{PhysRevA.86.013612,FortschrPhys.61.124}, the gain and loss can be interpreted as a coherent in- and outcoupling of particles, which may be possible to implement experimentally \cite{PhysRevA.90.042123,JPhysA.48.335302,PhysRevA.90.033630,PhysRevA.93.023624}. Therefore, Bose-Einstein condensates are a good candidate for the first quantum mechanical system to observe \PT symmetry.

In a number of works it has also been shown that even vortices are
found in nonlinear \PT-symmetric systems \cite{RevModPhys.88.035002,%
  PhysRevA.86.013808,Achilleos-solitons-ghosts}. However, most studies were
done for optical setups, where coupled wave guides forming a discrete
periodic structure were used \cite{RevModPhys.88.035002,OptLett.37.2148,%
  OptLett.38.371,OptLett.38.371,OptExpress.22.29679,OptLett.19.3177}.
It is the goal of this work to demonstrate that stable vortex excitations in
Bose-Einstein condensates with \PT-symmetric gain and loss can exist. We will
study the influence of the in- and outcoupling strength on stationary states
as well as on the dynamics of non-stationary states.
To do so we start with a vortex solution in a harmonic trap and
then investigate how it behaves if this stable situation is disturbed by a
current of particles. In contrast to systems with an azimuthal complex
potential, where the degeneracy between opposite directions of rotation is
lifted \cite{PhysRevLett.115.193902,OptLett.22.5194}, this is not the case
in our approach, where we are interested in a flow from one side of the trap
to the other. This models a vortex being formed in a trap of a larger
transport chain similar to the proposal presented in
\cite{PhysRevA.90.033630,PhysRevA.93.023624} for the generation of a
\PT-symmetric potential.

This paper is organized as follows. In Sec. \ref{sec:model} we introduce the model and explain the numerical techniques used. In Sec. \ref{sec:spectrum} we show and discuss two qualitatively different spectra containing vortices arising due to different imaginary potentials. Additionally, the transition from vortex to non-vortex states at a critical gain-loss parameter value and the influence of the interaction strength are explained. In Sec. \ref{sec:stability} we investigate the stability of states using the Bogoliubov-de Gennes formalism and a time evolution. In Sec. \ref{sec:dynamics} the dynamics of non-stationary vortex states in the presence of a \PT-symmetric potential is studied. Finally, in Sec. \ref{sec:conclusion} we summarize the obtained results.

\section{Model}\label{sec:model}
We consider a dilute Bose-Einstein condensate with contact interaction, which is described in the mean-field limit by the Gross-Pitaevskii equation. We assume that the condensate is confined in a harmonic oscillator trap,
\begin{align}
    V_{\mathrm T}(\vec r) = \frac m 2\left(\omega_0^2 (x^2 + y^2) + \omega_z^2 z^2\right)\,,
\end{align}
where $\omega_z \gg \omega_0$. Therefore, the condensate will be disc-shaped and the description can be effectively reduced to two dimensions \cite{JComputPhys.187.318,SIAMJMathAnal.37.189,ComputPhysComm.177.832,Nonlinearity.28.755}. In addition, an imaginary potential $V_{\mathrm I}$ is present, which describes the in- and outcoupling of particles. With the use of harmonic oscillator units,
\begin{align}
    \vec r &\rightarrow \frac{\vec r}{r_0}\,, &
    r_0 &= \sqrt{\frac{\hbar}{m\omega_0}}\,,\\
    \mu &\rightarrow \frac{\mu}{\mu_0}\,, &
    \mu_0 &= \frac{\hbar^2}{2mr_0^2}
\end{align}
the dimensionless effectively two-dimensional stationary Gross-Pitaevskii equation reads
\begin{align}
    \left[-\nabla^2 + V_{\mathrm T} + \im \Gamma V_{\mathrm I} + g|\psi|^2\right]\psi = \mu \psi\,,
\end{align}
with
\begin{align}
    \psi &= \psi(x,y)\,, & \iint |\psi(x,y)|^2\,\mathrm dx\,\mathrm dy &= 1\,,
\end{align}
and the effective two-dimensional chemical potential $\mu$. The trap potential reads in these units
\begin{align}
    V_{\mathrm T} &= V_{\mathrm T}(x,y) = x^2 + y^2\,,
\end{align}
and the interaction strength is $g = 8\pi N a/r_0$, where $a > 0$ is the effective two-dimensional repulsive s-wave scattering length.
All quantities used in this article are given in these dimensionless units.
The parameter $\Gamma$ determines the strength of the gain and loss.

\begin{figure}[t]
    \includegraphics[width=\columnwidth]{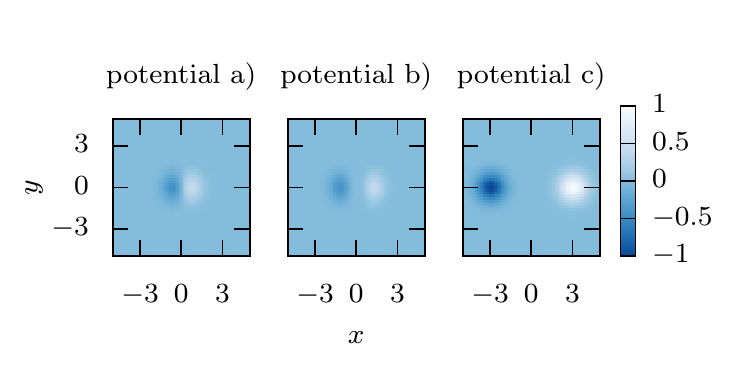}
    \caption{\label{fig:potentials}The three example potentials from Eq.\ \eqref{eq:VI} used in this work. For potential c) the parameter is $d=3$.}
\end{figure}

We investigate several different imaginary parts $V_{\mathrm I}$. Specifically, we will use three different potentials labeled a), b), c) to discuss the observed properties. They read
\begin{subequations}
\label{eq:VI}
\begin{align}
    V_{\mathrm I,a}(x,y) &= x\e^{-(x^2+y^2)}\,, \label{eq:VIa}\\
    V_{\mathrm I,b}(x,y) &= x^3\e^{-(x^2+y^2)}\,, \label{eq:VIb} \\
    V_{\mathrm I,c}(x,y) &= \e^{-(x-d)^2-y^2}-\e^{-(x+d)^2-y^2}\,, \label{eq:VIc}
\end{align}
\end{subequations}
and are shown in Fig.\ \ref{fig:potentials}. These potentials represent an incoupling on the right and an outcoupling on the left side. Thus, we expect a current from right to left.

To solve the Gross-Pitaevskii equation numerically, this partial differential equation was discretized and expressed in a harmonic oscillator basis, in which the problem reduces to a high-dimensional nonlinear root search. This works quite well as the potential is mainly harmonic oscillator-like and we are considering not too large nonlinearities. Here, a 2d-harmonic oscillator product basis up to the quantum numbers $n_{\mathrm{max}} = 11$ was used, which results in $N_{\mathrm{states}} = 78$ basis states. For the calculation the quadratic field of view was set to $x,y \in [-5,5]$ and discretized with $N = 128$ points.

We are interested in low-lying excited states, therefore initial guesses for the root search are the harmonic oscillator ground state, the first excited state oriented in $x$-direction as well as its counterpart in $y$-direction. A superposition of the two first excited states, one as real and one as imaginary part to imprint a vortex phase, is used as an initial guess for a vortex state.

\section{Stationary states and spectra}\label{sec:spectrum}
\begin{figure*}[t]
    \includegraphics[width=\textwidth]{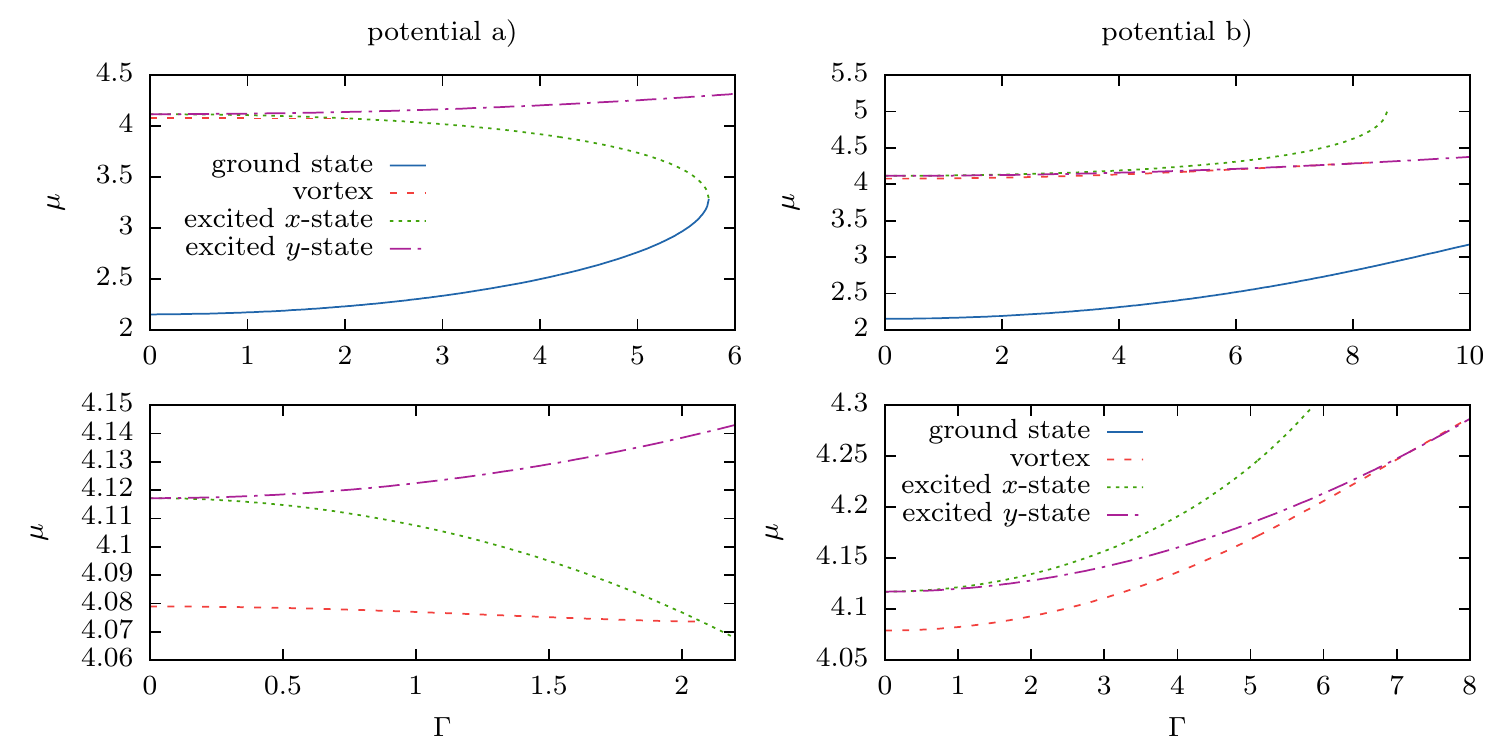}
    \caption{\label{fig:spectrum_g1}Spectra as a function of the gain-loss strength $\Gamma$ with the imaginary potentials \eqref{eq:VIa} (left) and \eqref{eq:VIb} (right) of the lowest states for $g=1$. The lower figures are extracts of the upper figures in the range of the bifurcation between the vortex and the excited state. The two degenerate vortex states are shown as a single state.}
\end{figure*}
First the interaction strength is fixed to $g=1$ and the parameter $\Gamma$ is varied to study the influence of the gain-loss effect. This results in two qualitative different spectra in dependence of the chosen imaginary part. Exemplarily for potentials \eqref{eq:VIa} and \eqref{eq:VIb} the spectra are shown in Fig.\ \ref{fig:spectrum_g1}.

What is common for all investigated potentials is that there is a range of
the gain-loss parameter $\Gamma$, in which a vortex state exists. Actually,
there are two vortex states which are degenerate as their only difference is
their direction of circulation. However, at a critical value of $\Gamma$ the
vortex merges with the lowest excited state in a pitchfork-bifurcation.
Note that this is a standard pitchfork bifurcation. The two vortices
with positive and negative circulation have identical stability properties.
Since they are on top of each other when the chemical potential is plotted
as in Fig. \ref{fig:spectrum_g1} they are only seen as one state. In the
following we restrict our discussion to the state with positive circulation.

It depends on the potential with which state the vortex merges. If the in- and outcoupling peaks lie close to the origin, the excited state oriented parallel to the direction of the external current (excited $x$-state) has lower chemical potential than the excited state oriented perpendicular to the external current (excited $y$-state). For larger distances from the origin the order of the energy values changes. In addition, the distance has an influence on a second tangent bifurcation of the excited state oriented in $x$-direction with the ground state. Only for lower distances of the gain-loss peaks from the origin this bifurcation exists, for larger values there is no such bifurcation.

\subsection{Influence of gain-loss peak position $d$}
\begin{figure}[t]
    \includegraphics[width=\columnwidth]{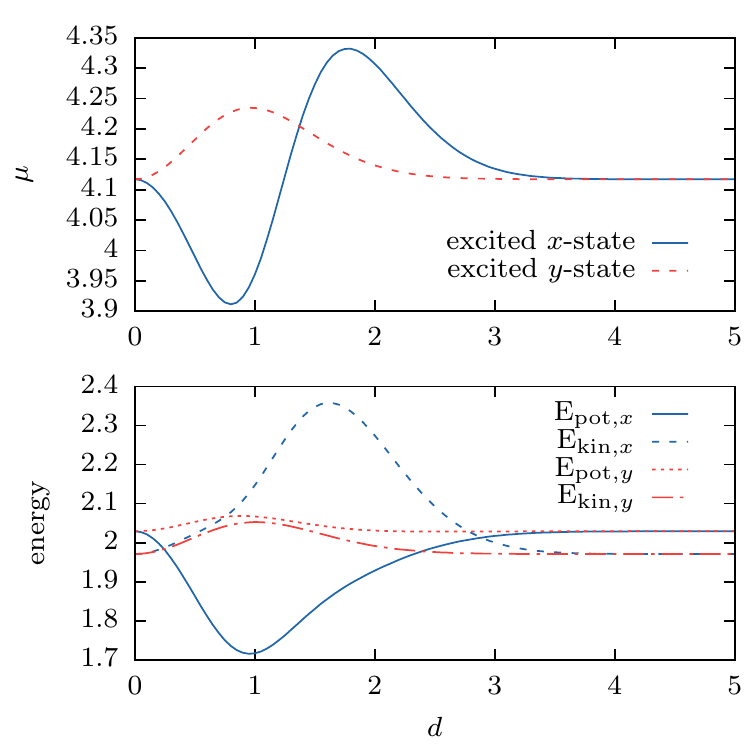}
    \caption{\label{fig:c_spectrum_d}Chemical potential (top) and the individual contributions E$_\mathrm{kin}$ and E$_{\mathrm{pot}}$ (bottom) of the excited states as a function of the gain-loss peaks position $d$ from the imaginary potential \eqref{eq:VIc}. The parameters are $g = 1$ and $\Gamma = 2$.}
\end{figure}

\begin{figure}[t]
    \includegraphics[width=\columnwidth]{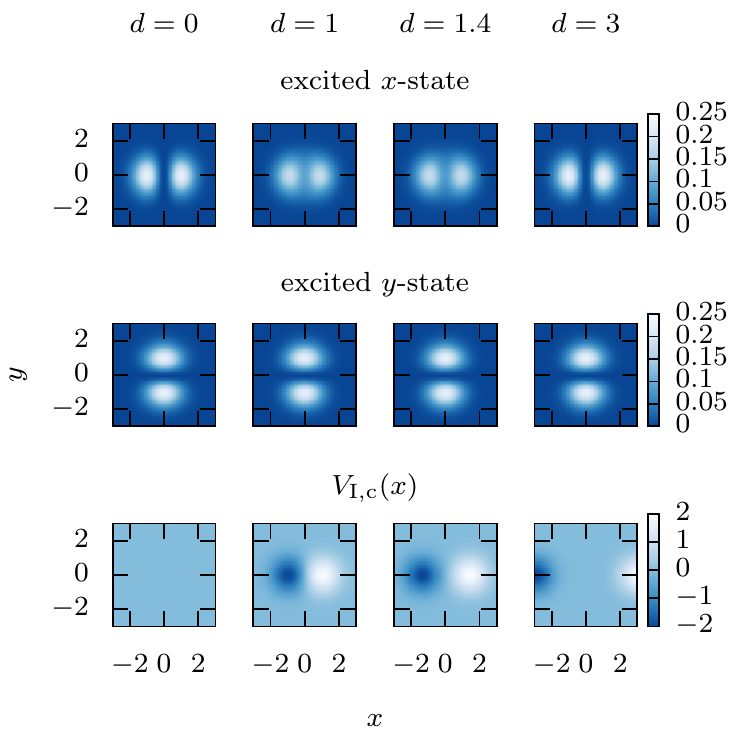}
    \caption{\label{fig:c_wave_function_d}Modulus squared of the wave functions of the excited states in potential \eqref{eq:VIc} (upper two rows) and the corresponding potentials (lower row) for different values of the gain-loss peaks position $d$. The parameters are $g = 1$ and $\Gamma = 2$.}
\end{figure}
Let us first discuss how the two qualitatively different spectra arise. For this analysis we use the potential \eqref{eq:VIc} with the variable distance $d$ of the gain-loss peaks. We fix the gain-loss strength to $\Gamma = 2$ and vary the distance $d$. In Fig.\ \ref{fig:c_spectrum_d} the chemical potential $\mu$ as well as the individual contributions of the average value of the kinetic and potential operator,
\begin{align}
    E_{\mathrm{kin}} &= \int |\nabla \psi|^2\,\mathrm d\vec r\,, & E_{\mathrm{pot}} &= \int V|\psi|^2\,\mathrm d\vec r\,,
\end{align}
are shown as functions of $d$, and in Fig.\ \ref{fig:c_wave_function_d} the excited states and the potential are shown for selected values of $d$. 

For this example the imaginary part vanishes for $d = 0$, as the in- and outcoupling peaks cancel each other. Therefore, the $x$- and $y$-states become degenerate. For increasing $d$ the in- and outcoupling peaks lie close to the origin at first. This causes a shift of the probability density of the $x$-state to the origin, which results in a decrease in the potential energy. As the gain-loss peaks are oriented in $x$-direction, the $y$-state does not experience such a shift and the potential energy does not drop.

At some value of $d$, the gain-loss peaks exceed the point at which the position of the $x$-state peaks without complex potential are located. Behind this point the $x$-state probability density is shifted outwards again, which leads to an increase of the potential energy. The $y$-state is again not substantially affected.

Eventually at a value of $d\approx 1.37$, the increase in the potential energy of the $x$-state leads to a higher chemical potential $\mu$ than that of the $y$-state. For even higher $d$ the peaks move out of the probability density range of the wave functions and the influence, i.e., the in- and outcoupling, vanishes. The chemical potentials of the two states approach each other again.

This explains also why the second bifurcation of the ground and an excited state only occurs for the excited $x$-state. The ground state resembles the excited $y$-state in $x$-direction and behaves similar to the excited $y$-state, i.e., it is also less affected by the gain-loss. Due to this similar behavior, the distance between their chemical potentials cannot change as much as between the excited $x$-state and the ground state, therefore, these states do not coalesce.

\subsection{Transition of vortex to non-vortex state}
\begin{figure*}[t]
    \includegraphics[width=\textwidth]{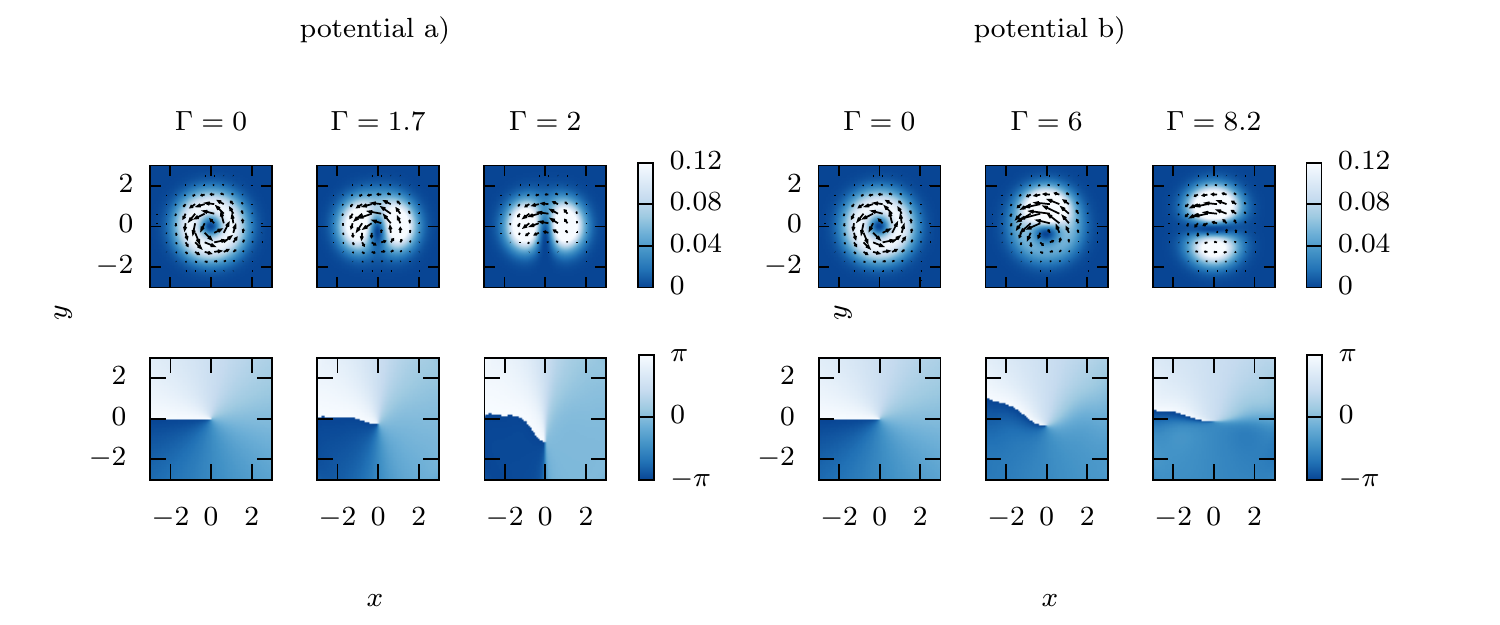}
    \caption{\label{fig:wave_functions_g1}Modulus squared of the wave functions with arrows representing the current density (top) and phase (bottom) for the imaginary potential \eqref{eq:VIa} (left) and \eqref{eq:VIb} (right) for different values of the gain-loss strength $\Gamma$. The nonlinearity is $g=1$.}
\end{figure*}
Now we discuss the transition of vortex to non-vortex states. The bifurcation point, at which the vortex changes to an excited state without vorticity, marks a so-called exceptional point \cite{kato-perturbation_theory_lin_op,JPhysA.45.444016,EurPhysJD.69.196,PhysRevA.93.123401,Nature.537.76,Nature.537.80}. At this point not only the energies but also the states themselves coalesce. This is a feature which cannot occur in Hermitian systems but can result in non-Hermitian or nonlinear Hamiltonians.

To explain this transition it is useful to consider the wave function and phase as well as the probability current density calculated with
\begin{align}
    \vec j = \im(\psi\nabla\psi^* - \psi^*\nabla\psi)\,.
    \label{eq:current_density}
\end{align}
These quantities are shown in Fig.\ \ref{fig:wave_functions_g1} for the potentials \eqref{eq:VIa} and \eqref{eq:VIb} for different values of  the gain-loss parameter $\Gamma$. For $\Gamma = 0$, i.e., no gain-loss, the vortex solutions are totally symmetric and positioned at the origin. Only the circular vortex current is present.

For increasing $\Gamma$ the behavior slightly differs in dependence of the imaginary potential. In the case of the bifurcation with the $x$-excited state (Fig.\ \ref{fig:wave_functions_g1} (left)), the vortex center shifts towards the bottom. For the (not shown) vortex with opposite circulation the center shifts towards the top. Due to the additional current from right to left, the shape of the vortex adapts itself in such a way that the probability density on the top is higher than that on the bottom by shifting the vortex center. Therefore, the natural vortex current is enhanced from right to left at the top, while the current is suppressed from left to right at the bottom. The basic shape of the wave function becomes more and more asymmetric in $y$-direction. The whole solution is cut into two parts. Finally, the vortex center is expelled in the region at the wave function's border, at which the probability density vanishes. At that point the vortex decays and changes into the first excited state with the two peaks on the left and right of the origin. The vortex current from left to right at the bottom comes to a halt as the external current is stronger than the vortex current.

\begin{figure}[t]
  \includegraphics[width=\columnwidth]{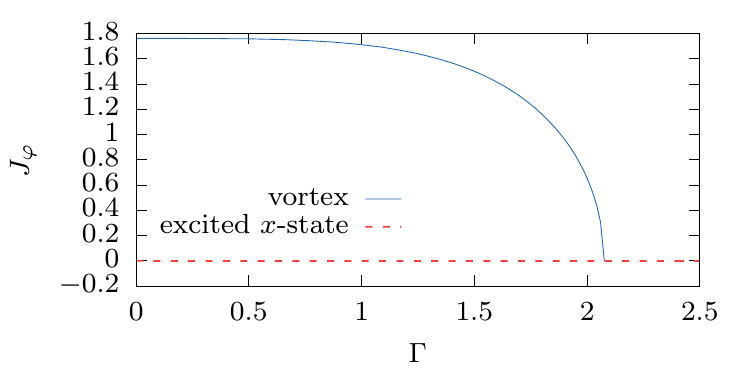}
  \caption{\label{fig:circular_current}Integrated projection $J_{\varphi}$ of
    the current density on the azimuthal direction for the
    vortex from Fig.~\ref{fig:wave_functions_g1}(a). It starts with a
    nonvanishing value for the vortex clearly distinguishing it from the
    excited $x$-state. Close to the bifurcation it rapidly decreases to
    zero.}
\end{figure}

This becomes also visible in Fig.~\ref{fig:circular_current}, in which the
integrated projection
\begin{equation}
  J_\varphi = \int \vec{j}\cdot\vec{e}_{\varphi} \,\mathrm d\vec r
  \label{eq:jphi}
\end{equation} 
of the current density on the azimuthal direction is shown as a function
of $\Gamma$. While this vanishes for the excited $x$-state it has a
nonvanishing value for the vortex from Fig.~\ref{fig:wave_functions_g1}(a).
For the vortex the value of $J_\varphi$ remains almost constant in a large
range of $\Gamma$. Close to the bifurcation it rapidly starts to decrease
and drops to zero at the bifurcation point confirming the transition from a
vortex to a non-vortex state.

The transition from the vortex to the excited state in $y$-direction (Fig.\ \ref{fig:wave_functions_g1} (right)) occurs in a different way. Here the vortex center performs only a slight movement in $y$-direction but then returns back to the center. An ejection in $x$-direction, which would be the nodal line for the excited $y$-state, was not possible since a movement of the vortex center in this direction would break the \PT symmetry of the wave function. What happens is that the probability density nearly drops to zero in the lower half of the vortex, in which the current flows in opposite direction to the external current. This brings the circular vortex current to a halt and the vortex dies at its location. Thus, the vortex does not vanish as a result of the center ejection but simply by stopping the current in one direction while the center stays on the nodal line of the excited $y$-state.

\subsection{Influence of the interaction strength $g$}
After having fixed $g=1$ in the previous examples, now the influence of higher interaction strengths will be discussed. What can we expect? As we are considering repulsive interaction, a higher strength will spread the wave function more to the outside. We have shown that the type of spectrum we obtain depends on the position $d$ of the in- and outcoupling peaks relative to the natural wave function position without gain and loss. Therefore, at a small fixed distance $d$ an increase of $g$ should not change the spectra qualitatively as the in- and outcoupling peaks are located inside the wave function's maxima anyway. In the case of the second spectra from Fig.\ \ref{fig:spectrum_g1} (right side) for large $d$, a qualitative transition to the type of the first spectra may occur as the probability density of the wave function can be shifted beyond the in- and outcoupling peaks. However, this is something which happens not in every case and depends on the actual form of the imaginary part. 

\begin{figure*}[t]
    \includegraphics[width=\textwidth]{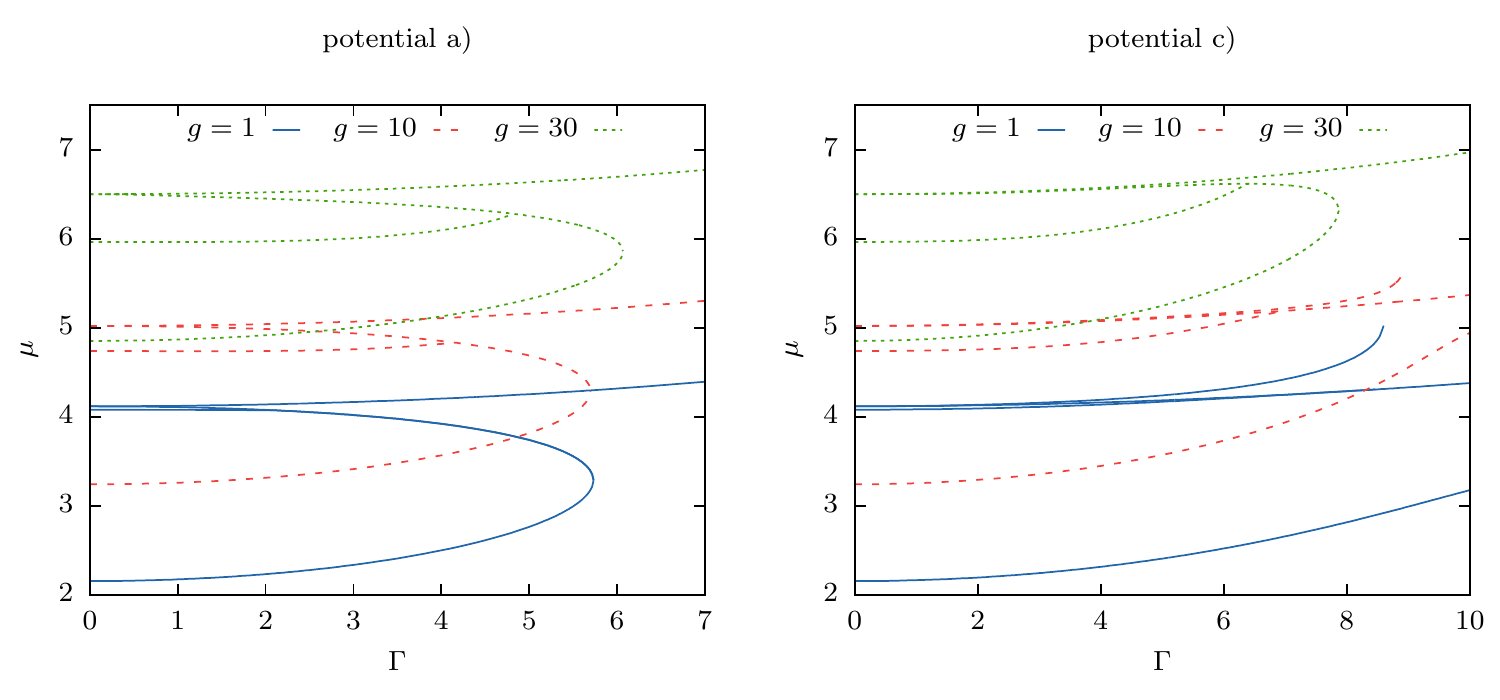}
    \caption{\label{fig:spectrum_g}Spectrum in dependence of the gain-loss strength $\Gamma$ of the lowest states for potential \eqref{eq:VIa} (left) and \eqref{eq:VIb} (right) for different values of the nonlinearity $g$. For potential \eqref{eq:VIa}, the spectrum stays qualitatively the same independent of $g$. The spectrum for potential \eqref{eq:VIc} changes qualitatively to the spectrum for potential \eqref{eq:VIa} for increasing $g$.}
\end{figure*}

For the example potentials \eqref{eq:VIa} and \eqref{eq:VIb} the spectra for different values of $g$ can be found in Fig.\ \ref{fig:spectrum_g}. In both cases the separation between vortex and excited state increases for increasing nonlinearity. This results from the fact that, due to the two strongly located peaks of the excited states in comparison to the more blurred vortex, the interaction energy of the excited states increases stronger for higher interaction strength. Therefore, the vortex states become energetically more favorable for increasing $g$.

The spectra for potential \eqref{eq:VIa} stay qualitatively the same, while for potential \eqref{eq:VIb} a transition of the spectra type can be observed. However, as already mentioned, this is nothing what happens in general. Calculations with different potentials do not always show such a transition. To be precise, in these cases the excited $x$-state may be located below the excited $y$-state in some regions but eventually the positions switch again.

\section{Stability analysis}\label{sec:stability}
In reality the system cannot be perfectly isolated and there will always be small fluctuations. Only if the states survive within such an environment and live for a sufficiently long time, they may be observed experimentally. Therefore, it is important to check the stability of the states.

First the linear stability is checked using the Bogoliubov-de Gennes formalism. Within this formalism, a small time-dependent fluctuation $\Delta(\vec r,t)$ is added to a stationary state $\psi_0(\vec r)$
\begin{align}
    \psi(\vec r,t) = \left[\psi_0(\vec r) + \Delta(\vec r,t)\right]\e^{-\im \mu t}\,,
\end{align}
where one uses the ansatz
\begin{align}
    \Delta = \epsilon\left(u(\vec r)\e^{-\im\omega t} + v^*(\vec r)\e^{\im\omega^* t}\right)
\end{align}
with a positive value $\epsilon \ll 1$. Inserting this ansatz into the time-dependent Gross-Pitaevskii equation and neglecting terms in the order of $\mathcal O(\epsilon^2)$ one obtains the linear Bogoliubov-de Gennes equations, which can be written in matrix form as
\begin{align}
    \begin{pmatrix}
        A & B \\
        -B^* & -A^*
    \end{pmatrix}
    \begin{pmatrix}
        u \\ v
    \end{pmatrix} = \omega
    \begin{pmatrix}
        u \\ v
    \end{pmatrix}
\end{align}
with
\begin{align}
    A &= -\nabla^2 + V_{\mathrm T} + \im\Gamma V_{\mathrm I} + 2g|\psi_0|^2 -\mu\,,\\
    B &= g\psi_0^2\,.
\end{align}
This linear eigenvalue problem can again be expressed and diagonalized in a harmonic oscillator basis. If there are solutions with complex frequencies $\omega$, the fluctuations will exponentially increase, which is a clear sign of an unstable state. Due to the \PT symmetry, if $\omega$ is a solution, then $-\omega$, $\omega^*$ and $-\omega^*$ are also solutions, therefore, it is sufficient to consider only $\omega$ with positive imaginary part.

\begin{figure}[t]
    \includegraphics[width=\columnwidth]{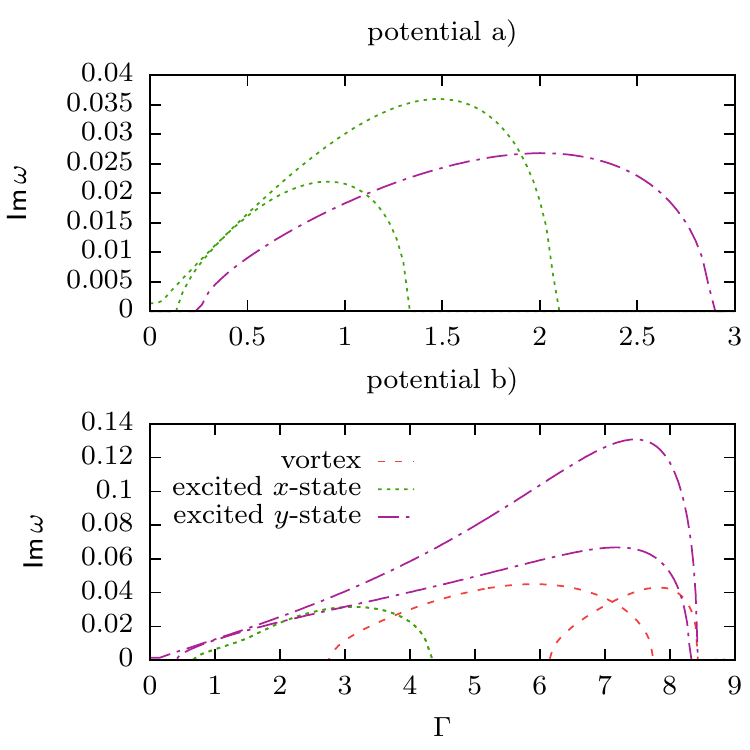}
    \caption{\label{fig:bdg_g1}Positive imaginary parts of the Bogoliubov-de Gennes eigenvalues $\omega$ for potential \eqref{eq:VIa} (top) and \eqref{eq:VIb} (bottom). The nonlinearity is $g=1$. The ground state is always stable and has no imaginary contribution.}
\end{figure}

In Fig.\ \ref{fig:bdg_g1} the imaginary eigenvalues are shown for the two different spectra for the potentials \eqref{eq:VIa} and \eqref{eq:VIb}. In the first case, the ground and the vortex state are always stable. The excited $x$-state is unstable in the range where it coexists with the vortex but becomes stable after the bifurcation point. In the second case the vortex is not stable over the whole range in which it exists. Only for smaller values of $\Gamma$ it is stable, at a value of $\Gamma \approx 2.7$ there is an unstable region until the end of the vortex existence.

To consider also nonlinear fluctuations and confirm the Bogoliubov-de Gennes analysis, a time evolution is performed. We take the exact eigenstates, add random noise and let it evolve in time. If the state is stable, it will only perform small fluctuations or slightly oscillate but keep its basic shape over time, whereas an unstable state will completely deform or decay.

The time evolution is performed using the split-operator method, where the time evolution operator $U = \exp(-\im H t)$ is split symmetrically for a small time step $\mathrm dt$ as
\begin{align}
    U(\mathrm dt) \approx \e^{-\im \frac{-\nabla^2}{2} \mathrm dt}\e^{-i \tilde V \mathrm dt} \e^{-\im \frac{-\nabla^2}{2} \mathrm dt} + \mathcal O(\mathrm dt^3)
\end{align}
with
\begin{align}
    \tilde V = V_{\mathrm T} + \im\Gamma V_{\mathrm I} + g|\psi_0|^2\,.
\end{align}
This expression is evaluated with forward and backward Fourier transforms
\begin{multline}
    \psi(\mathrm dt) = U(\mathrm dt)\psi_0 \\
    \approx \mathcal F^{-1} \left[ \e^{-\im \frac{|\vec k|^2}{2} \mathrm dt} \mathcal F \left[\e^{-i \tilde V \mathrm dt} \mathcal F^{-1} \left[\e^{-\im \frac{|\vec k|^2}{2} \mathrm dt} \mathcal F[\psi_0] \right] \right]\right]\,.
\end{multline}
As we work with a finite grid, reflections at the border or wraparounds in the Fourier space can occur. These artifacts can be reduced by increasing the discretization range or introducing absorbing borders \cite{JComputPhys.63.363}. However, in this case it is not necessary to remove all the artifacts, as they can be considered as additional distortions. This confirms the stability of states even more if they survive under this condition. In our case, only for potential \eqref{eq:VIb} for values of $\Gamma \gtrapprox 2$ the discretization range has to be doubled to $x,y \in [-10,10]$ and $N=256$ grid points in each direction to allow for the required calculations.

\begin{figure}[t]
    \includegraphics[width=\columnwidth]{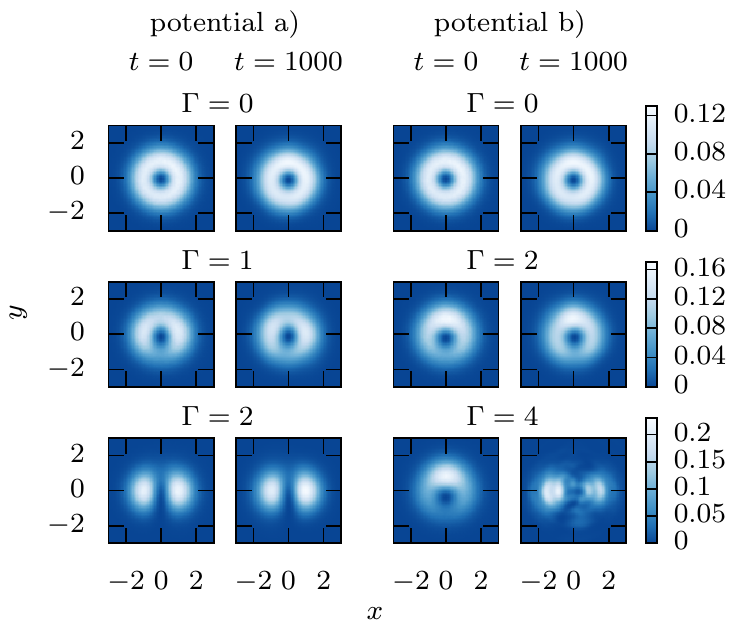}
    \caption{\label{fig:stability_time_evolution_g1}Time evolution for 1000 time steps of the stationary vortex state of potential \eqref{eq:VIa} (left) and \eqref{eq:VIb} (right) for $g=1$ and different values of $\Gamma$. The initial state is distorted with random noise on the order of $10^{-2}$ and propagated with the split-operator method with a step size of $\mathrm dt = 10^{-3}$.}
\end{figure}

In Fig.\ \ref{fig:stability_time_evolution_g1} the time evolution of the distorted vortex states is depicted for different values of the gain-loss parameter $\Gamma$. In each case random data on the order of $10^{-2}$ is added to the initial state. The time evolution is performed over 1000 time steps with a step size of $\mathrm dt = 10^{-3}$. The distorted initial state and the final state after the time evolution are shown. This time evolution confirms the result from the Bogoliubov-de Gennes analysis. For the potential \eqref{eq:VIa} the vortex state is stable independently of $\Gamma$, whereas for potential \eqref{eq:VIb} the state is stable only for smaller values of $\Gamma$ and gets finally unstable.

\section{Dynamics}\label{sec:dynamics}
Lastly, we study the dynamics of non-stationary vortex states in a \PT-symmetric potential. We ask how a vortex reacts and evolves in time in presence of a balanced net current through the condensate.

Vortex movement has already been discussed in literature \cite{JPhysCondensMatter.13.R135,JPhysB.36.3467,OptCommun.152.198,PhysRevA.61.063612} and there are several facts known. The motion of a Bose-Einstein condensate with its center of mass positioned out of the trap center can be described like that of a classical single particle in an effective potential \cite{PhysRevA.66.033610,JPhysB.36.3467,PhysRevA.66.053608}. The movement of the particle corresponds to the center of mass movement of the condensate. A movement of the vortex relatively to its background causes a Magnus force on the vortex perpendicular to its velocity \cite{PhysRevLett.85.2857}, which results in a precession around the trap center. The Magnus force has a hydrodynamic cause created by a pressure gradient resulting from a suppression and an enhancement of the vortex current on the opposite sites.

\begin{figure*}
    \includegraphics[width=\textwidth]{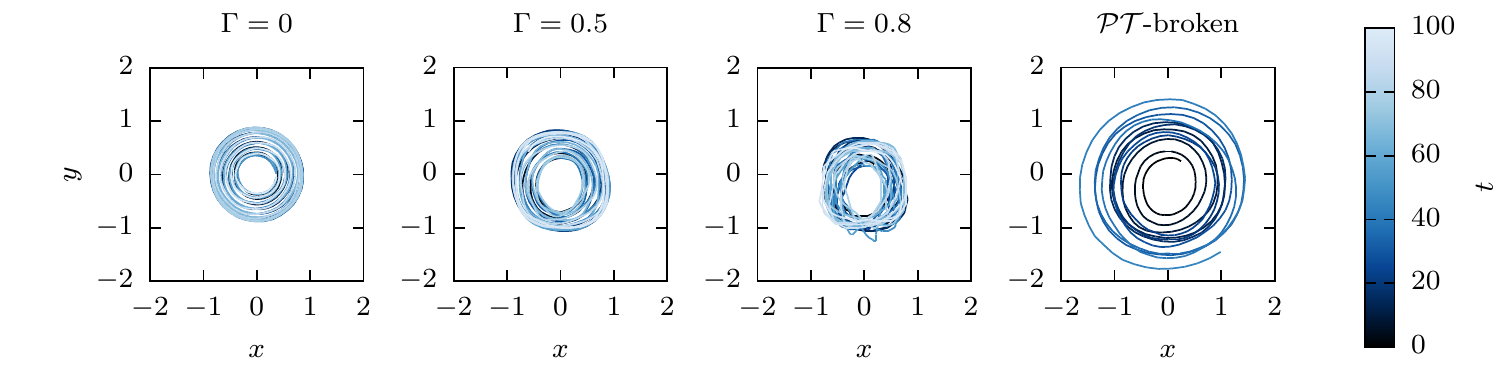}
    \caption{\label{fig:dynamics}Vortex dynamics of the non-stationary state \eqref{eq:dynamics_init} with $x_0 = y_0 = 0.2$. The first three plots are results for potential \eqref{eq:VIc} with $d=1$ and $\Gamma = 0$, $\Gamma = 0.5$ and $\Gamma = 0.8$. The fourth plot is the result for the \PT-broken potential \eqref{eq:VI_ptbroken} with $\Gamma = 0.5$. In each plot the vortex center trajectory is shown as a function of time. The time evolution was performed with the split-operator method with a step size of $t = 10^{-3}$.}
\end{figure*}

For our study we use potential \eqref{eq:VIc} with $d=1$. First we set the gain-loss parameter $\Gamma = 0$ and place a vortex initially at an off-center position via
\begin{align}
    \psi &= \left[(x-x_0)+\im (y-y_0)\right]\e^{-\frac 1 2 (x^2 + y^2)}\label{eq:dynamics_init}
\end{align}
and let it propagate in time. The position of the vortex center is determined via a parabolic fit at the minimum. The resulting trajectory can be found in the first plot of Fig.\ \ref{fig:dynamics}. The expected circular motion around the trap center can clearly be observed.

Now we turn on the external current via our \PT-symmetric complex potential, i.e., $\Gamma \neq 0$. The resulting trajectory for two different values of $\Gamma$ can be seen in the middle plots of Fig.\ \ref{fig:dynamics} with the same initial vortex as in the previous case. Despite the additional current, one can see that a stable movement is still possible. The circular trajectory is slightly distorted which results from the fact that our initial symmetric vortex is a bad approximation of the asymmetric stationary vortex. This results in deformation of the background cloud and an additional movement of the vortex.

There is a second feature, which is not so obvious at first glance. The precession of the vortex is no longer symmetric regarding the origin, but with respect to a point slightly below. The whole trajectory is shifted. This can be understood by remembering the stationary solutions calculated in section \ref{sec:stability}. There it was found that the eigenstates are also no longer symmetric in $y$-direction but the vortex center is shifted. Because of this, the equilibrium point is no longer the origin but a point below it. Consequently, the precession occurs around this new equilibrium point.

The \PT-symmetric gain-loss is crucial for a stable movement. If we explicitly break the \PT-symmetry, e.g., by increasing the strength of the incoupling slightly,
\begin{align}
    V_{\mathrm I}(x,y) &= 1.2\e^{-(x-d)^2-y^2} - \e^{-(x-d)^2-y^2}\label{eq:VI_ptbroken}
\end{align}
two things can be observed. On the one hand the background grows as more particles are inserted than removed. On the other hand the trajectory becomes unstable, which can be seen in the last plot in Fig.\ \ref{fig:dynamics}.
The vortex spirals out until it hits the cloud border and vanishes. A \PT-broken gain and loss has a destabilizing effect on the movement and the trajectory becomes unstable.

\section{Conclusion}\label{sec:conclusion}
In this paper we demonstrated that stationary vortex states in Bose-Einstein condensates with balanced gain and loss of particles can exist. Independently of the actual form of the imaginary potential we found always a range of the in- and outcoupling strength $\Gamma$ which allows for the existence of stable vortices. If this parameter exceeds a critical value, i.e., the point at which the additionally introduced external current is stronger than the internal vortex current, the vortex decays. In the presence of the external current the stationary vortex adapts itself in such a way that it shifts its vortex center and changes the background shape such that a net current exists. If the vortex decays it changes into the energetically closest excited state, which manifests itself in a bifurcation and an exceptional point in the spectra.

A time evolution of non-stationary vortex states in presence of \PT-symmetric potentials showed that the external current shifts the precession center of the vortex movement. This center corresponds to the position of the stationary vortex center and is the new equilibrium point of the dynamics. A balanced gain and loss is crucial for a stable movement as unbalanced gain and loss leads to an unstable dynamics shown through an outspiraling and decay of the vortex.

\end{document}